\documentclass[showpacs,aps,preprintnumbers,prb,amsmath,amssymb,twocolumn,floatfix,superscriptaddress]{revtex4-2}
\usepackage{graphicx}
\usepackage{hyperref}
\usepackage{color}
\usepackage{bm}
\usepackage{amssymb}
\usepackage{amsmath}
\usepackage[normalem]{ulem}
\usepackage[T1]{fontenc}
\usepackage{multirow}
\usepackage{xcolor}
\usepackage{soul}
\usepackage{hyperref}
\hypersetup{
    colorlinks=true,
    linkcolor=blue,
    filecolor=blue,      
    urlcolor=blue,
    citecolor=blue}

\newcommand{\lc}{\lowercase}
\begin{document}
\urlstyle{same}

\title{Griffiths phases in structurally disordered CeRhSn: Experimental evidence and theoretical modeling}
\author{Andrzej~\'{S}lebarski}
\affiliation{\href{https://ror.org/04xdyc983}{Institute of Low Temperature and Structure Research}, Polish Academy of Sciences, Ok\'{o}lna 2, 50-422 Wroc{\l}aw, Poland}

\author{Maciej~M.~Ma\'ska} 
\affiliation{Institute of Theoretical Physics, \href{https://ror.org/008fyn775}{Wroc\l aw University of Science and Technology}, Wybrze\.{z}e Wyspia\'{n}skiego 27, 50-370 Wroc\l aw, Poland}

\begin{abstract}
Our report paves the way for insight into a structural disorder and its impact on the physical properties of strongly correlated electron systems (SCESs). In a critical regime, each perturbation, e.g., disorder due to structural defects or doping, can have a significant effect on the nature of the quantum macrostate of these materials. For a select group of SCESs, we have empirically documented the Griffiths singularity, as exemplified by CeRhSn, which exhibits non-Fermi-liquid characteristics in susceptibility and specific heat. Our numerical analysis has supported the Griffiths phase scenario for CeRhSn and has revealed that its dc magnetic susceptibility is strongly dependent on the size of inhomogeneous magnetic particles that form in these materials. In the presence of strong disorder, we have proposed a magnetic phase diagram for CeRhSn. 
The {\it classical Griffiths phase} has been identified in the temperature range below the onset temperature of $T_G\sim 220$ K, while the {\it quantum Griffiths phase} with non-Fermi liquid behavior emerges below the quantum critical temperature of $T_Q\sim 6$ K. The phase diagram developed in this study bears notable similarities to the scenario previously proposed by Vojta for magnetic quantum phase transitions in disordered metals.
\end{abstract}


\maketitle
\section{Introduction}

The theory of non-interacting electrons in perfect metallic crystals does not apply directly to the class of disordered and correlated electronic systems. This simple approach falls short of describing condensed matter systems of interest today, where the Coulomb interaction between electronic states determines the low-temperature  behaviors of the metal. 
These materials, even when obtained as good quality single crystalline samples, are never free of disorder, and emergent properties arising from the impact of defects and interactions could often be clearly seen. 
The pioneering work of Anderson \cite{Anderson1958} led to an understanding of how an electron can stop diffusing and become localized when a crystal is sufficiently disordered. Thus, disorder can be a reason for metal-insulator transitions. The situation is much more complicated in the presence of Coulomb correlations. The effect of atomic as well as thermal disorder on the electronic properties of strongly correlated electronic systems (SCESs) has been the cause of intense later research, both experimental \cite{Mydosh1999} and theoretical \cite{Spalek2003,Grenzebach2008}. Of particular interest are systems that are close to a quantum critical point, where the non-Fermi liquid (NFL) \cite{JLTPh1996} and the Mott-Hubbard localization \cite{Gebhard1998} are observed. In such a case, the interplay of disorder and interatomic interactions emerges as a pivotal issue, especially in the critical regime of the system being at the threshold of instability. In this regime, structural defects, acting as perturbations, can produce a disproportionate effect by changing the nature of the quantum macrostate of the system (cf. Ref. \cite{Spalek2003}). 

A route of NFL behavior that emphasizes a disorder-driven mechanism, known as {\it Kondo disorder}, has previously been reported in Refs. \cite{Dobrosavlevic1992,Bernal1995,Miranda1996,Miranda1997}.
The examples of strongly correlated disordered systems comprise such materials as various heavy fermion alloys: UCu$_{5-x}$Pd$_x$ \cite{Andraka1993,Bernal1995},  $M_{1-x}$U$_x$Pd$_3$ ($M=$ Sc, Y) \cite{Seaman1993,Maple1995}, La$_{1-x}$Ce$_x$Cu$_{2.2}$Si$_2$ \cite{Andraka1994a}, Ce$_{1-x}$Th$_x$RhSb \cite{Andraka1994b}, and U$_{1-x}$Th$_x$Pd$_2$Al$_3$ \cite{Maple1995}, to name a few, which exhibit non-Fermi-liquid features, and  are  not magnetically ordered.
As theoretically analyzed by Miranda {\it et al.} \cite{Miranda1996,Miranda1997}, the effects of disorder in these Kondo alloys on their low-temperature physical properties have been documented, including the distribution of local Kondo temperatures and the subsequent breakdown of the conventional Fermi liquid behavior.

Another scenario for the NFL behavior has been proposed by Castro Neto {\it et al.}  \cite{Castro1998,Castro2000}, who have attributed the strong temperature dependence observed in low temperature specific heat $C(T)$  and magnetic susceptibility $\chi(T)$ of many chemically substituted f-electron systems (cf. \cite{Andrade1998,Volmer2000}) to the presence of the Griffiths phase (GP) state \cite{Griffiths1961,comment1}.
Castro Neto {\it et al.}  have shown that, close to the quantum critical point (QCP), strong responses from quantum fluctuation of magnetic clusters give singular contributions to the temperature dependence of the physical properties in these materials. In a finite region around the QCP (not just at criticality) defined as the {\it quantum Griffiths phase}, thermodynamic observables are expected to be singular. These singularities are caused by the presence of the so-called {\it rare regions}. Since the disorder is random, there is an (exponentially small) chance that some regions will have fluctuations so small that a local ``long-range'' order can form. The larger the region, the smaller the chance that this occurs. Despite this, these regions have been shown to still affect the system near the critical point by enhancing the thermodynamic response \cite{Vojta2010}. In particular,
this temperature region features power laws, e.g., in specific heat, $C/T \sim T^{\lambda-1}$, magnetic susceptibility, $\chi\sim T^{\lambda-1}$, and magnetization isotherms, $M\sim B^{\lambda}$, and is characterized by the Griffiths exponent $\lambda<1$.
The {\it rare regions} are locally ferromagnetic within the Griffiths phase state, even though the bulk system is globally in the paramagnetic phase \cite{Vojta2010}. Vojta \cite{Vojta2010} proposed a phase diagram of the Griffiths phase region as a function of temperature $T$ and the quantum tuning parameter $p$. In this scenario, the strong disorder may result in the emergence of an {\it infinite-randomness quantum critical point} in a cluster glass area. Concurrently, a quantum Griffiths phase and NFL behavior manifest over an extended range above quantum criticality, giving rise to the power-law singularities of thermodynamic observables.
The number of new compounds that exhibit Griffiths-McCoy singularities continues to increase.

The aim of this work is the characterization of the Kondo lattice CeRhSn with short-range ferromagnetic correlations, which lead to the formation of the Griffiths phase state. 
We mostly focus on magnetic susceptibility (ac, dc) at various magnetic fields to propose modeling of the magnetic clusters formed in the Griffiths phase state below the onset temperature $T_G$.
We consider a polycrystalline system to emphasize the disorder-related effects, which in this case are presumably more pronounced.

The paper is organized as follows: In Sec. \ref{sec:kondo-previous} a brief review of previous studies on the NFL behavior of CeRhSn is provided. Section \ref{sec:experiment} describes the experiments performed. In Sec. \ref{sec:results} an analysis of the obtained data is presented in relation to {\it ab initio} calculations. The results are then supported by a toy model in Sec. \ref{sec:model}. The work is concluded in Sec. \ref{sec:conclusions}.

\section{Kondo lattice C\lc{e}R\lc{h}S\lc{n}; previous study\label{sec:kondo-previous}}    

We have previously shown \cite{Slebarski2002a,Slebarski2002b,Ho2004} that CeRhSn exhibits NFL temperature dependencies in its low-temperature physical properties and is located near the QCP. Our research on polycrystalline samples documented electrical resistivity $\Delta \rho (T) \equiv \rho(\mathrm{CeRhSn}) - \rho(\mathrm{LaRhSn}) \propto T$, similar power laws in magnetic susceptibility and electronic specific heat coefficient, $\chi\sim C/T\sim T^{-n}$ with $n=1-\lambda \cong 0.5$, and magnetization $M\sim B^{\lambda}$. These power-law singularities have suggested the quantum Griffiths phase state for CeRhSn.  Later, similar power laws for $\chi$, $C/T$, and $M$ were experimentally confirmed for the single crystalline CeRhSn sample \cite{Kim2003}.
Since then, CeRhSn has been systematically studied, mainly with the aim of interpreting its quantum criticality \cite{Kim2003,Tou2004,Tokiwa2015,Kittaka2021,Kimura2024,Bohm2024}. Very similar power laws were experimentally confirmed for a single-crystalline CeRhSn sample \cite{Kim2003}.

The first measurement of the linear thermal expansion coefficient $\alpha$ divided by temperature $T$, $\alpha(T)/T$, did not confirm its divergence down to 2 K for the polycrystalline CeRhSn sample \cite{Slebarski2004}. However, it has been found that the critical Gr\"{u}neisen ratio $\Gamma^\mathrm{cr}\propto \beta^\mathrm{cr}/C^\mathrm{cr}$ shows the $1/T^{\epsilon}$ behavior between 2 and 8 K with exponent $\epsilon\cong 1.2$~\cite{Slebarski2011a}, where $\beta^\mathrm{cr}$ and $C^\mathrm{cr}$ are the thermal volume expansion and the specific heat with the background contributions subtracted (for the polycrystalline sample $\beta=3\alpha$ was assumed). The singularity of $\Gamma^\mathrm{cr}$ in a finite region around the QCP suggests the presence of the quantum GP. We also note that $\epsilon=1$ is expected for the 3D antiferromagnetic QCP \cite{Zhu2003}.   

Recent measurements of heat capacity and linear thermal expansion down to millikelvin temperatures have indicated the NFL ground state for CeRhSn \cite{Tokiwa2015} with a clear divergence found in $\Gamma(T)$ for $T<0.5$ K. 
Moreover, a pronounced anisotropy of the linear thermal expansion coefficient divided by temperature, $\alpha/T$, was found. Specifically, $\alpha_c/T$ measured along the $c$ axis was temperature independent, in contrast to the divergence found in $\alpha_a/T$ along $a$.
This anisotropy in $\alpha/T$ has been attributed to a quantum critical state induced by geometric frustration in planes perpendicular to the $c$ axis \cite{comment_QCP}. In the case of CeRhSn with hexagonal ZrNiAl-type structure, its Ce atoms indeed form a quasi-kagome lattice \cite{Tokiwa2015} with possible frustrated interactions \cite{Kittaka2021}. The presence of local Ce moments was postulated based on the metamagnetic crossover observed at the lowest temperatures \cite{Kittaka2021} (note that, at $T=1.5$ K any magnetic order within a limit for an ordered moment of 0.25 $\mu_B$, has not been detected, using a high-resolution neutron spectrometer \cite{Slebarski2002a}). 
The scenario of frustration-induced quantum criticality appears to be correct for describing the thermodynamic properties of CeRhSn at the lowest temperatures (low energy limit) and does not exclude the presence of the Griffiths phase state, which has been explicitly documented experimentally at much higher temperatures.

\section{Experiment\label{sec:experiment}}

\begin{figure}[h!]
\includegraphics[width=0.48\textwidth]{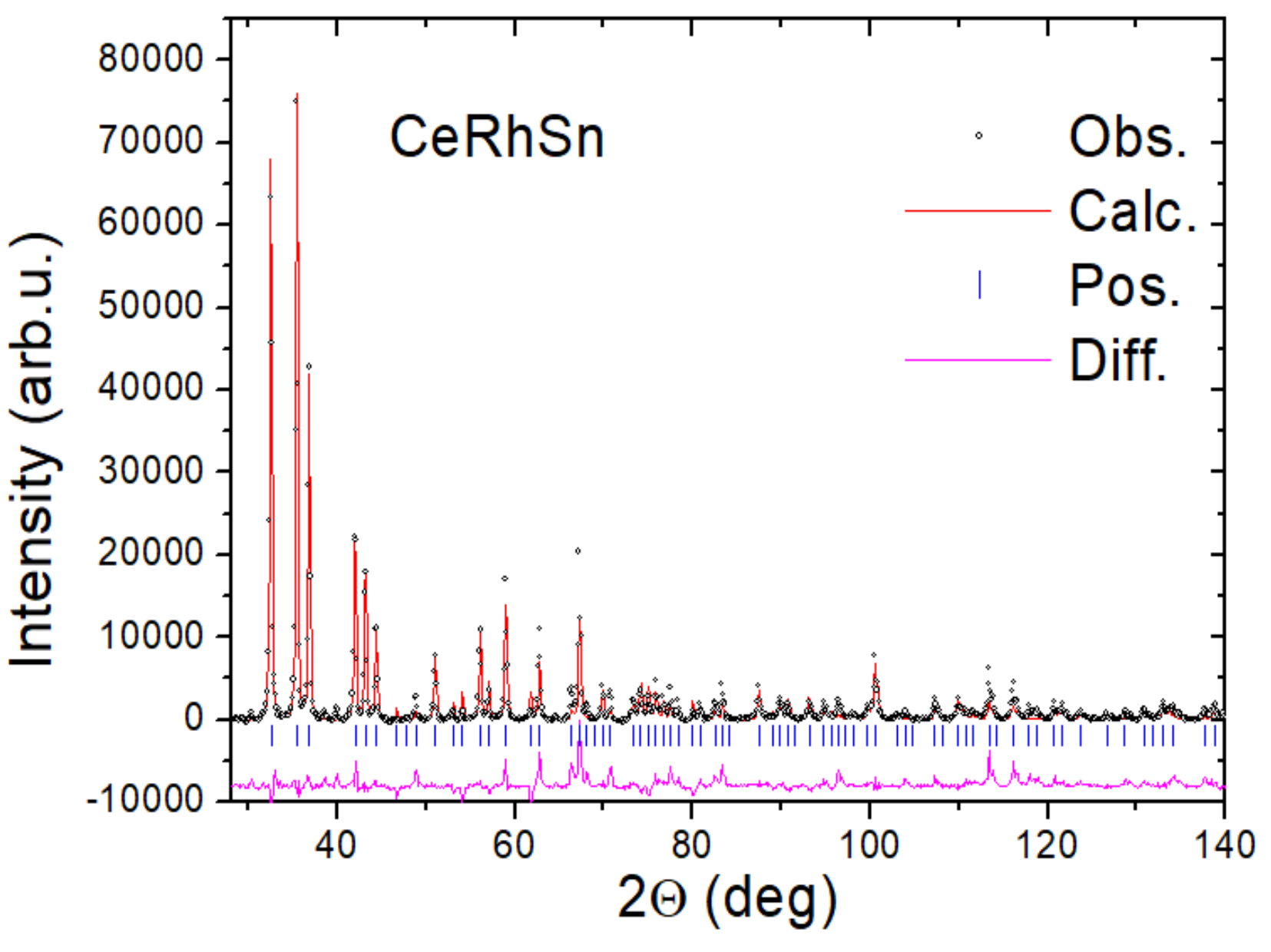}
\caption{\label{fig:XRD}
Observed (black points) and calculated (red line) profiles for CeRhSn at RT (the intensity is expressed in arbitrary units). The short vertical lines indicate the positions of Bragg reflections for the parameters listed in Table \ref{tab:TableXRD}. The difference between observed and calculated profiles is shown at the bottom of the figure. The lack of a perfect agreement between the measured and calculated XRD spectra is attributed  to atomic disorder in the sample.
}
\end{figure}

Polycrystalline samples of CeRhSn and LaRhSn were prepared by the arc melting technique and subsequent annealing  at 800$^{\circ}$C for 2 weeks.
The samples were examined by x-ray diffraction (XRD) analysis (PANalytical Empyrean diffractometer equipped with a Cu K$\alpha_{1,2}$ source) and found to be a single-phase with hexagonal structure of the Fe$_2$P-type and space group $P\bar{6}2m$.
The XRD patterns were analyzed with the Rietveld refinement method using the FULLPROFFsuite set of programs \cite{Rodriguez1993}. 
The results of the room temperature (RT) refinements are presented in Table \ref{tab:TableXRD}.
\begin{table}[h!]
\caption{Lattice parameters $a$ and $c$ at RT. The results were obtained for each sample with the weighted-profile $R$ factors \cite{Toby2006} $R_\mathrm{exp}<0.9$\%, $R_{wp}<4.5$\%, and $R_\mathrm{Bragg}<2.9$\%.
}
\label{tab:TableXRD}
\begin{tabular}{ccc}
\hline\hline
compound & $a$ (\AA) & $c$ (\AA) \\
\hline
CeRhSn &  7.4486 & 4.0807\\        
LaRhSn & 7.4911 & 4.2231\\    
\hline\hline
\end{tabular} 
\end{table}
Figure \ref{fig:XRD} displays the observed and calculated XRD profiles in RT for CeRhSn. 

The stoichiometry and homogeneity of the samples were checked using an electron energy dispersive spectroscopy technique. 
The atomic percentage of the specific element content in the obtained samples is very close to the assumed composition. Microprobe measurements revealed the chemical compositions that correspond to the formulas Ce$_{0.99}$RhSn$_{0.99}$ and La$_{0.98}$RhSn$_{0.99}$, obtained respectively with accuracy better than 1.5\% for Ce and La, and better than 1\% for Rh and Sn, and thus confirm the desired stoichiometry.
\begin{figure}[h!]
\includegraphics[width=0.42\textwidth]{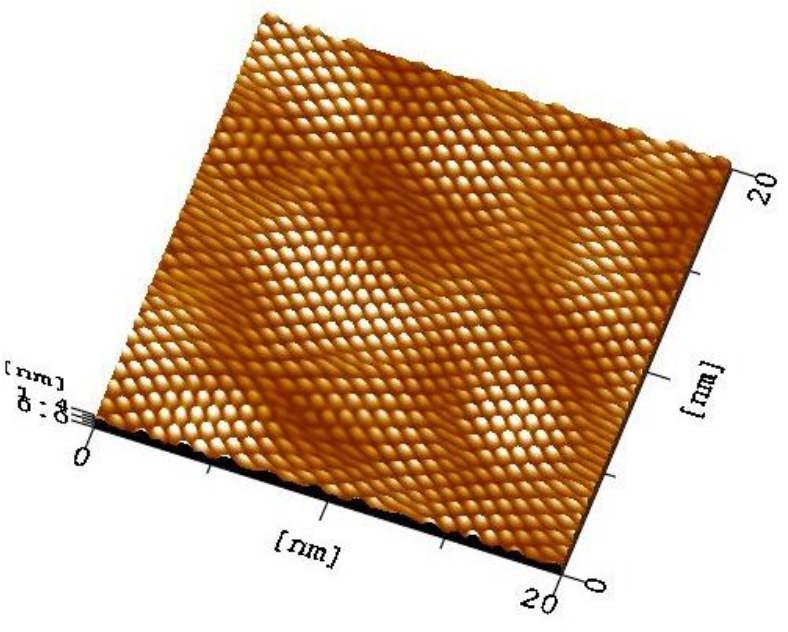}
\caption{\label{fig:STM}
CeRhSn; STM image from a small part of the surface with atomic resolution.
}
\end{figure} 

The scanning tunneling microscopy (STM) experiment allowed us to determine the real structure of the sample. 
This is a good idea for experimentally confirming a tendency for clustering of atoms.
Microscopic analysis of CeRhSn with atomic resolution using a JEOL JSPM-4210A AFM setup has revealed nanometer-sized crystalline grains separated by intercrystalline regions with point defects and dislocations, related to the unit cell of the sample, as shown in Fig. \ref{fig:STM}. 
The high-resolution STM image obtained at room temperature and at high vacuum shows the topographical map of the flat surface of the sample, which initially was formed using a diamond saw and mechanically polished. 
The sample was annealed at 300$^\circ$C under vacuum in the microscope chamber (in situ) immediately before measurement; this thermal treatment allowed partial desorption of water and other hydroxyl groups from the sample surface.
The aim to present this surface  image  was to show that the CeRhSn system is disordered. Further, the structural disorder is the cause of the formation of magnetic moments, which create dilute clusters. We  write about this extensively in Sec. IV.B., it was also documented in our previous DFT calculations (Sec. IV.A.).

The ac magnetic susceptibility was measured in the temperature range 2-300 K with an ac field of 2 Oe for various frequencies using a Quantum Design PPMS platform. The dc magnetic measurements were performed in the temperature interval $1.8 - 300$ K and magnetic fields up to 1 T using a Quantum Design PPMS platform equipped with a vibrating sample magnetometer option.  

The heat capacity was measured in the temperature range $1.8-300$ K and in external magnetic fields up to 14 T using the same PPMS platform. Apiezon N cryogenic high vacuum grease product \cite{Apiezon} was used to mount the CeRhSn and / or LaRhSn samples on the sapphire sample holder platform to ensure good thermal contact from the heater to the sample. However, it should be noted that the heat capacity of Apiezon N has a measurable contribution to the addenda \cite{Schnelle1999}, which was subtracted from the raw specific heat data.

\section{Results and discussion: Griffiths phase and its evolution with atomic disorder\label{sec:results}}
\subsection{Predictions of the Griffiths phase state in CeRhSn: A possible quantum Griffiths region with the power-law singularities of susceptibility and electron specific-heat coefficient}

\begin{figure}[h!]
\includegraphics[width=0.45\textwidth]{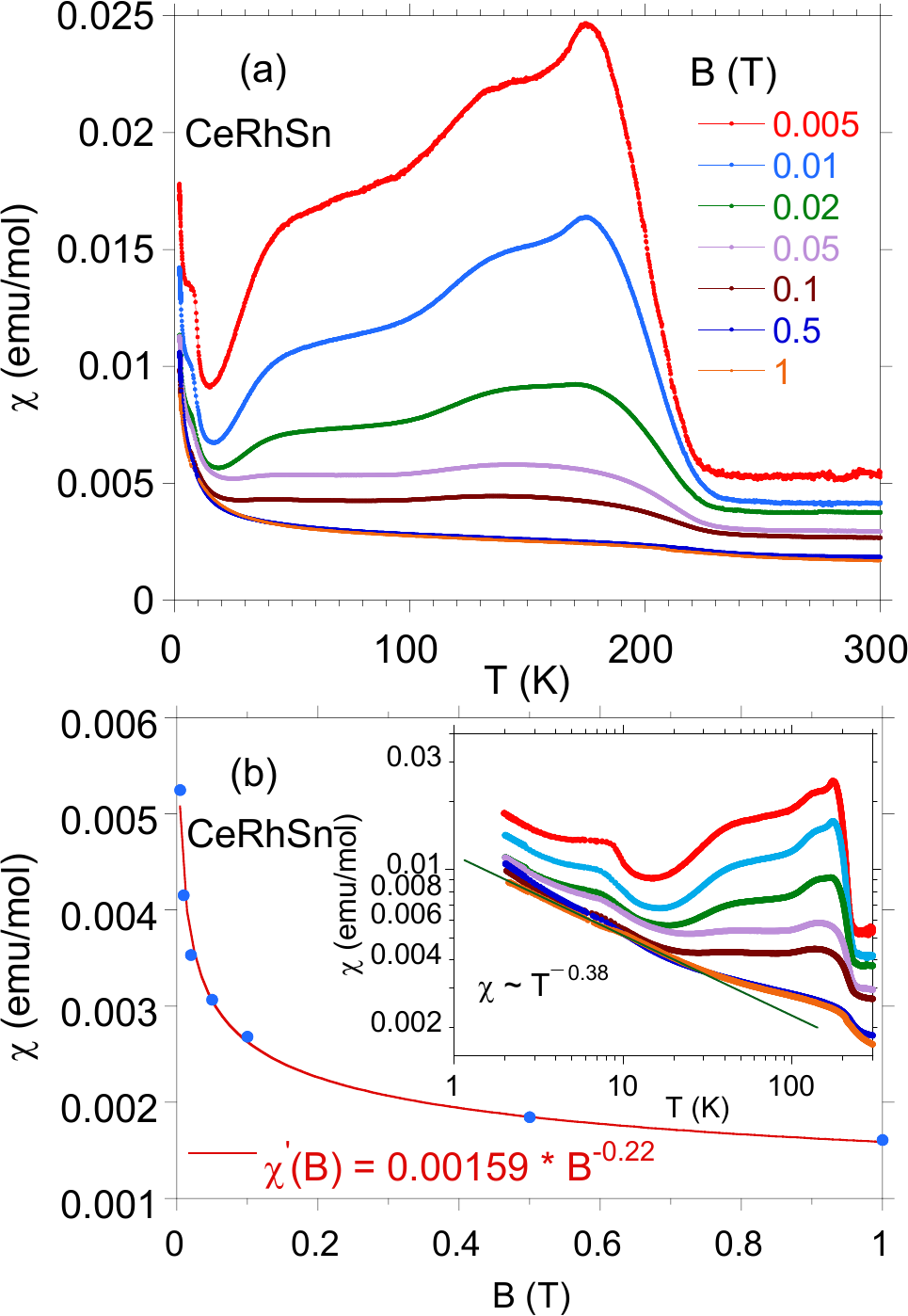}
\caption{\label{fig:Fig_CeRhSn_chi-H_A-B}
(a) The dc magnetic susceptibility vs. temperature data measured for CeRhSn in different values of applied fields, measured in
field-cooled mode.
(b) $\chi$ (dc)  at 300 K vs magnetic field (blue points), and power-law fit to the data (red line). The inset exhibits $\chi(T)$ in log-log scale and $\chi\sim T^{-0.38}$ approximation (green line).
}
\end{figure}

\begin{figure}[h!]
\includegraphics[width=0.45\textwidth]{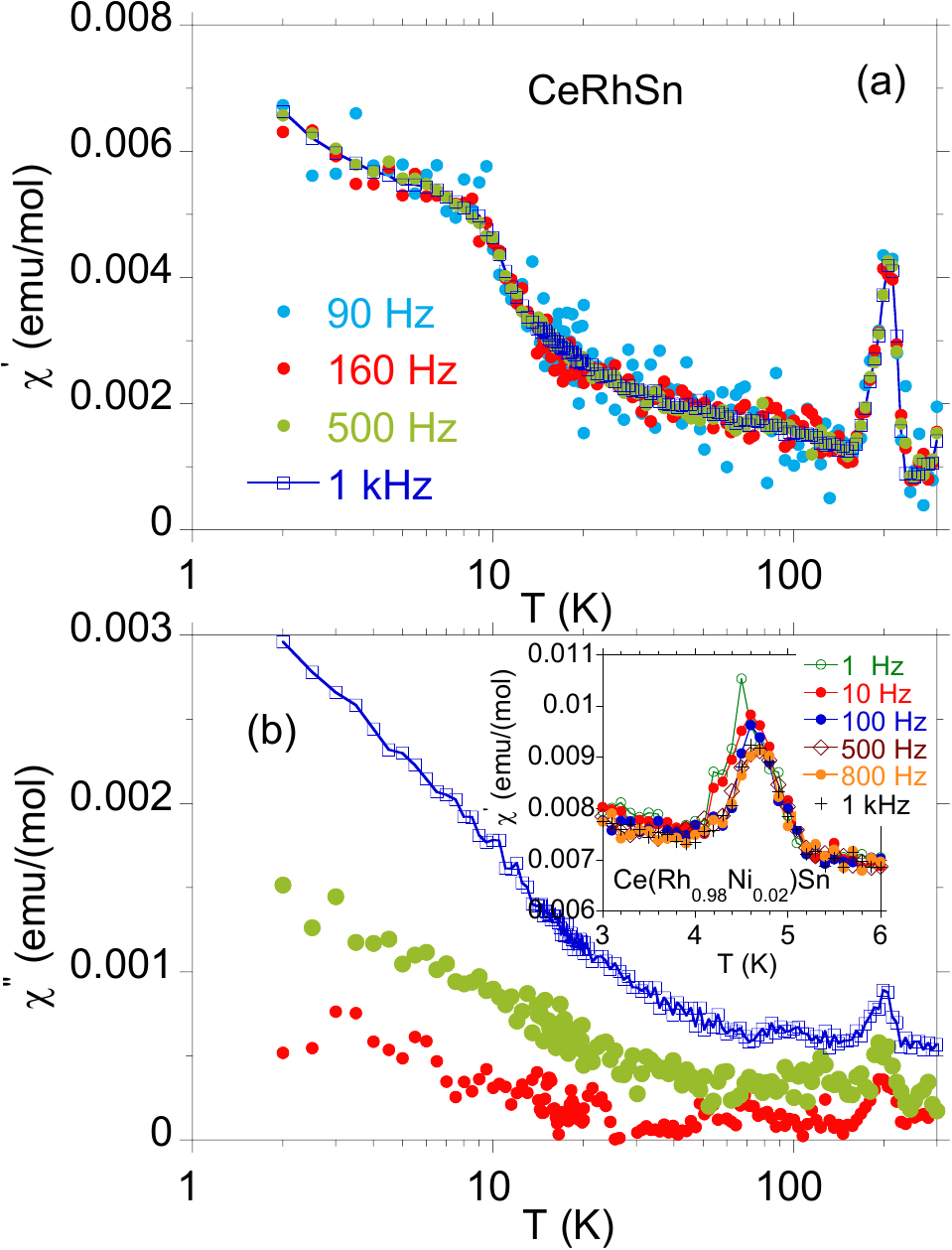}
\caption{\label{fig:CeRhSn_CHI_ac}
Temperature (log scale) dependence of the real (a) and imaginary (b) part of the ac magnetic susceptibility  for CeRhSn measured for various frequencies of the applied magnetic field; the amplitude of the magnetic field was 2 G. The inset shows  $\chi\:'$ vs $\nu$ for CeRh$_{0.98}$Ni$_{0.02}$Sn sample with enhanced disorder by doping of CeRhSn with Ni. The $\nu$-effect is weak and can be detected for $\nu<100$ Hz.
}
\end{figure}

The Fe$_2$P (ZrNiAl) type crystal structure of CeRhSn  has a planar geometry with the spacing  of about 3.9~\AA~between Ce ions in the $ab$ plane (002), larger than the Hill limit \cite{Hill1970}, while the Ce-Ce distances are much larger in the direction perpendicular to this plane.  Consequently, direct $f-f$ overlap can be negligible, while the hybridization mainly between the Ce $4f$ and Rh $4d$ states, significantly enhanced by disorder, could be the reason for the {\it weakly magnetic} properties of CeRhSn.
Indeed, our previous numerical calculations of the density of electronic states (DOS) for the Ce-Rh-Sn alloys with assumed various local environments of Ce atoms yielded a magnetic moment $\mu\neq 0$ per formula unit only for the disordered alloy with one ($\mu =0.25\:\mu_B$) or two ($\mu =0.34\:\mu_B$) Rh atoms occupying Ce positions, respectively \cite{Slebarski2003}. In contrast, the structurally ordered system was calculated to be paramagnetic.
To confirm the {\it ab initio} predictions, the magnetic susceptibility of dc and ac was measured as a function of temperature in various magnetic fields or at various frequencies, respectively.
As shown in Figs. \ref{fig:Fig_CeRhSn_chi-H_A-B} and \ref{fig:CeRhSn_CHI_ac}, the presence of an incomplete structural ordering results in the formation of a weakly magnetic phase below $T_G\approx 220$ K, which bears  similarity to the Griffiths phase scenario.
The Griffiths phase consists of magnetic clusters in a paramagnetic region at $T \gg T_C$ as a result of the competition between the Kondo and Ruderman-Kittel-Kasuya-Yosida (RKKY) interactions in the presence of structural disorder \cite{commentK}. Magnetic susceptibility shown in Fig. \ref{fig:Fig_CeRhSn_chi-H_A-B} does not follow the Curie-Weiss law for $T<T_G$  and is strongly field dependent, 
while within the low-temperature region ($T<7$ K) $\chi\sim T^{-0.38}$ ($B=1$ T), as shown in the inset of Fig. \ref{fig:Fig_CeRhSn_chi-H_A-B}(b), and magnetization isotherm at 1.8 K $M\sim B^{0.65}$.   
Figure \ref{fig:Fig_CeRhSn_chi-H_A-B}  shows a rapid decrease in $\chi(T)$ between $T_{CG}\sim 35$ K and $\sim 15$ K due to cluster stiffening that leads to the formation of diluted cluster spin glass (CG). $T_{CG}$ is the freezing temperature of the diluted clusters at which the formation of the cluster spin glass-like state begins.
This CG state could be interpreted in terms of {\it rare regions} \cite{Vojta2010}, where the slow dynamics of clusters can give rise to the power-law singularities of the observables studied. These singularities have been observed in the experimental data at the lowest temperatures [see the inset of Fig. \ref{fig:Fig_CeRhSn_chi-H_A-B}b]. 
In this temperature range the magnetic clusters are strongly diluted and large, thereby rendering the ac susceptibility practically independent of the frequency $\nu$ of the applied magnetic field when $\nu>100$ Hz, as illustrated in Fig. \ref{fig:CeRhSn_CHI_ac}. The weakly detectable dependence of $\chi\:'$ on $\nu$ can only be observed for the sample with disorder enhanced by doping at frequencies lower than 100 Hz, as shown in the inset in Fig. \ref{fig:CeRhSn_CHI_ac}(b).

We postulate that $T_Q\approx 6$ K separates the {\it classical Griffiths phase} and the {\it quantum Griffiths phase} with non-Fermi liquid behavior (cf. Ref. \cite{Vojta2010}). The power-law singularities in  $\chi(T)$ could  be modified at much lower temperatures by a possible metamagnetic transition, as suggested in Ref. \cite{Tokiwa2015}.
In Sec.~\ref{sec:model} we propose a simple model that relates the increase of $\chi$ below $T_G$ to the formation of magnetic clusters.

The progressive decrease of $\chi$ (dc) with increasing applied field shown in Fig. \ref{fig:Fig_CeRhSn_chi-H_A-B} is characteristic of the classical GP state and allowed us to distinguish the Griffiths singularity from the smeared phase transition between the paramagnetic and ferromagnetic states.
As shown in Fig.~\ref{fig:Fig_CeRhSn_chi-H_A-B}(b), this dependence of $\chi$ on $B$ is well approximated by the power law. However, the field effect in $\chi(T)$ becomes negligible for fields $B>0.5$ T, indicating the presence of maximally large magnetic clusters (cf. Fig.~\ref{fig:Fig_CeRhSn_chi-H_A-B}).
A similar cluster-size effect stimulated by a magnetic field has already been experimentally suggested \cite{Cox1994,Costro1997,Jo2006} for a variety of nanoparticles. In Sec. \ref{sec:model}, we propose a theoretical explanation for the growth of the magnetic clusters with increasing field $B$.
\begin{figure}[h!]
\includegraphics[width=0.42\textwidth]{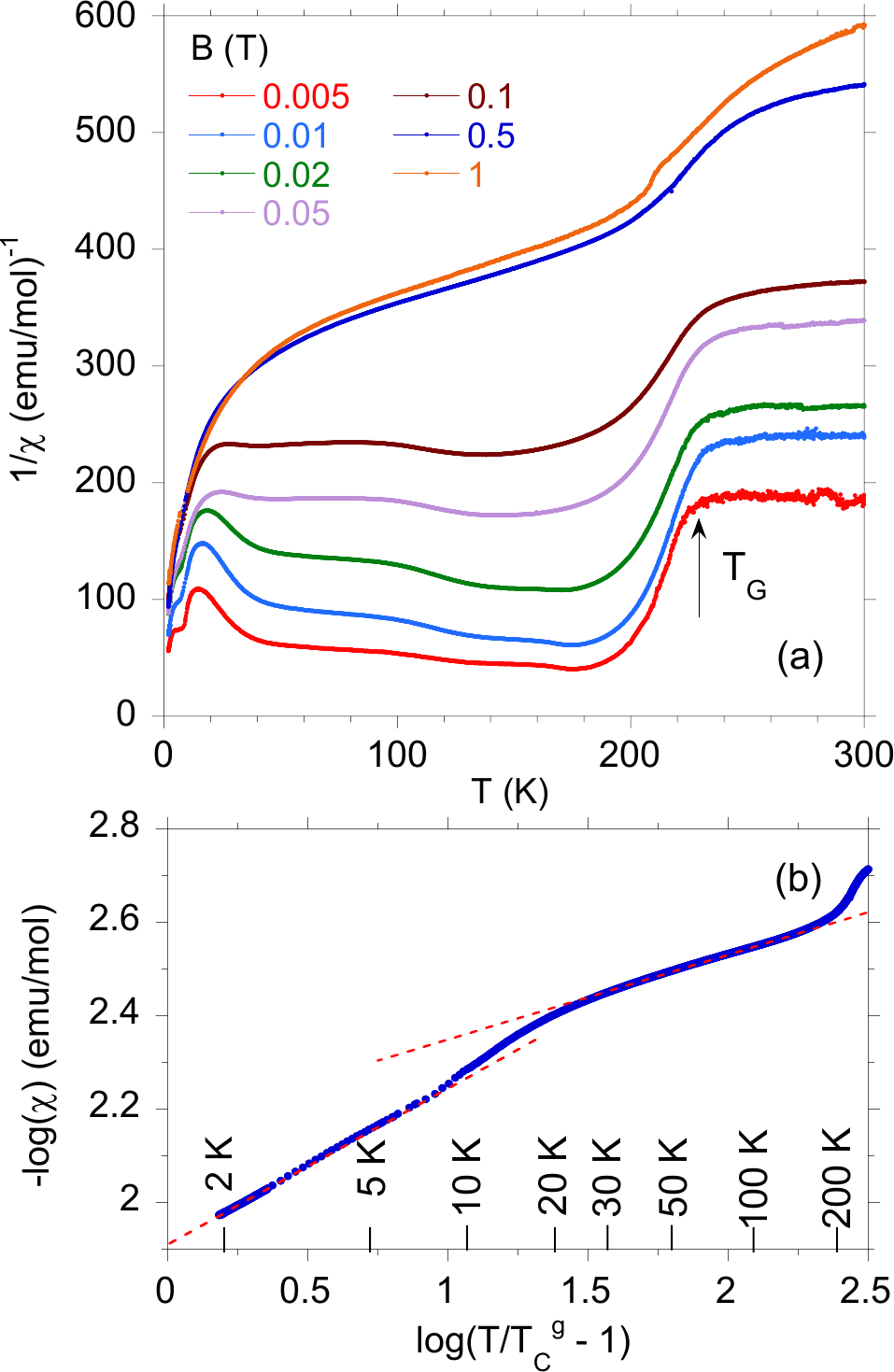}
\caption{\label{fig:CeRhSn_inverse_CHI}
(a) Temperature dependence of the inverse dc susceptibility $\chi^{-1}$ under different fields for CeRhSn. (b) $1/\chi$ under magnetic field $B=0.5$ T vs. reduced $T$, $(T/T_C^g-1)$ in log-log scale.
}
\end{figure}  

Figure \ref{fig:CeRhSn_inverse_CHI} shows the inverse dc susceptibility, $\chi^{-1}$, as a function of $T$ for CeRhSn. Since CeRhSn is a valence fluctuation-type paramagnet \cite{Slebarski2002a}, in the conventional paramagnetic region it does not follow the Curie law and $\chi^{-1}$ shows a tendency to saturation with increasing $T$, as shown in Fig. \ref{fig:CeRhSn_inverse_CHI}(a) within $T>T_G$. Below $T_G$ $\chi^{-1}$ displays a {\it downward deviation} that strongly depends on the field. This behavior is considered a hallmark of Griffiths singularity. The softening of the {\it downward behavior} in $\chi^{-1}$ and progressive increase of $\chi^{-1}$ in the field are typical characteristics of the GP state, thereby justifying the Griffiths singularity scenario, in contrast to the typical magnetic phase transition between the ferromagnetic and paramagnetic state.

Within the Griffiths model, the GP singularity is expressed by a power law in the  inverse susceptibility, $\chi^{-1}\propto (T-T_C^g)^{(1-\lambda)}$ \cite{Castro1998}, where $0<\lambda < 1$ and $T_C^g$ is the critical temperature of random ferromagnetism of the sample where $\chi$ tents to diverge. Figure \ref{fig:CeRhSn_inverse_CHI}(b) shows a change of $\log(\chi^{-1})$ depending on $\log(T/T_C^g-1)$, which for $T<T_G$ can be well described by the expression $\chi^{-1}\propto (T-T_C^g)^{(1-\lambda)}$ for $T<T_G$ with the fitting parameters $T_C^g=0.8$ K and $\lambda=0.45$. 
However, we can note two linear regimes with different slopes of $\log(\chi^{-1})$, as highlighted by red dashed lines. The first linear behavior is clearly visible between $T_G$ and $\sim 30$ K, while the second is observed for temperatures $T<T_Q$. We attribute the slope effect to the presence of quantum and classical Griffiths phase states,  separated by $T_Q$. Furthermore, we also suppose that the {\it slope-effect} shown in Fig. \ref{fig:CeRhSn_inverse_CHI}(b) may be observed in a number of other similar SCESs. In such a case, this behavior could serve as a useful criterion for distinguishing between the various Griffiths phase states in the {\it classical} and {\it quantum} regions.

\begin{figure}[h!]
\includegraphics[width=0.45\textwidth]{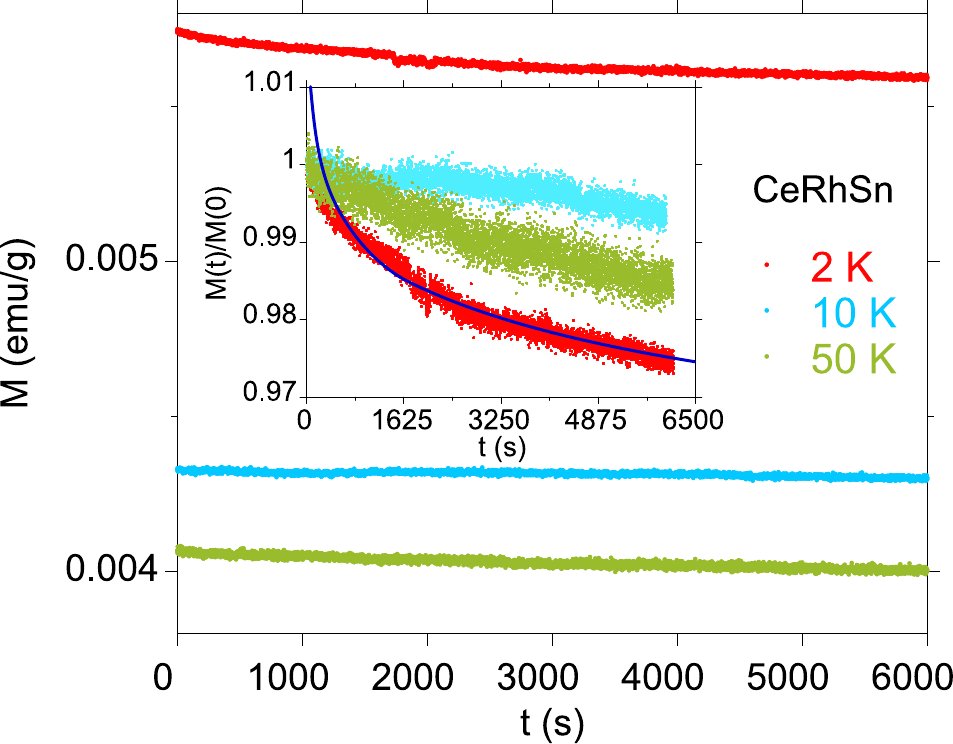} 
\caption{\label{fig:CeRhSn_M-t}
Time dependent remnant magnetization $M$ for CeRhSn at various temperatures. 
The inset shows $M(t)$ normalized to the $M(t=0)$. The blue  solid line represents a fit to 
equation  $M(t)= M(0)\propto t^{-\alpha}$ for the data at $T=2$ K  with the fitting parameter
$\alpha = 7.9 \times 10^{-3}$.
}
\end{figure}
The isothermal magnetic relaxation phenomenon in CeRhSn was investigated as a final test of the formation of the glassy state in this compound. The sample was first zero field cooled from 300 K down to 2 K with a constant cooling rate and kept at a target temperature for a waiting time $t_w=300$ s in a field of 5000 G. Then, the field was switched off. Figure \ref{fig:CeRhSn_M-t} displays the time evolution of magnetization, $M(t)$, measured in zero-field-cooled mode at temperatures 2, 10, and 50 K with an applied field of 0.11 G. 
The observed time dependence of isothermal remnant magnetization $M(t)$ at $T=2$ K is $t$ dependent and can be fitted by a power law, $M(t)=M(0)t^{-\alpha}$, with the fitting parameter $\alpha=7.9\times 10^{-3}$. 
This type of time dependence of the magnetization is expected for CeRhSn at temperatures $T_C^g<T<T_{CG}$. Above $T_{CG}$ the power-law behavior of $M(t)$ is not observed, as shown in Fig. \ref{fig:CeRhSn_M-t}.

\begin{figure}[h!]
\includegraphics[width=0.42\textwidth]{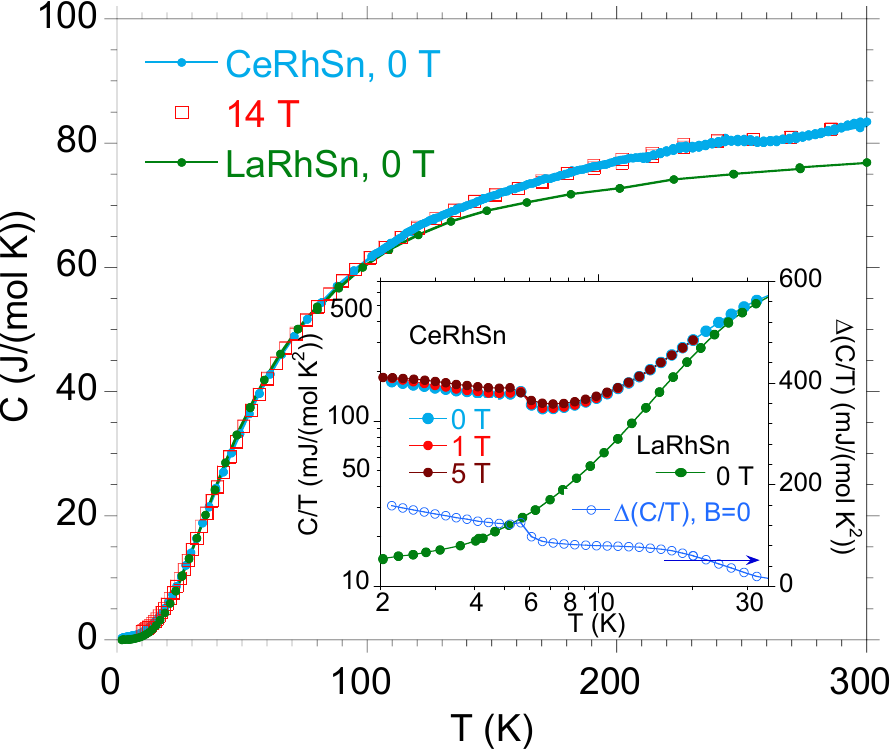}
\caption{\label{fig:Fig_C-CeRhSn}
Temperature dependence of the specific heat $C$ for CeRhSn at the magnetic field $B=0$ and 14 T in comparison with $C$ of LaRhSn at $B=0$. The inset shows the low-temperature $C(T)/T$  data obtained for CeRhSn at the field $B=0$, 1, and 5 T in comparison with $C(T)/T$ for LaRhSn ($B=0$). $\Delta (C/T)$ exhibits a difference between $C/T$ of CeRhSn and LaRhSn.
}
\end{figure}

Figure \ref{fig:Fig_C-CeRhSn} displays the specific heat $C$ of CeRhSn and its reference LaRhSn compound at various magnetic fields. The heat capacity, $C(T)$, of CeRhSn exhibits some notable feature at $T\sim T_G$, which is not field dependent and is not typical of the magnetic phase transition, and has not been detected for LaRhSn. This feature could be attributed to the appearance of the GP state in CeRhSn, especially because the high-temperature specific heat anomaly well correlates with the thermal expansion anomaly at $\sim 220$ K, a finding that had previously been demonstrated for CeRhSn through x-ray diffraction measurements (cf. Ref. \cite{Slebarski2002c}).  At the same time, one also needs to take into account that the contribution of Apiezon N to the raw CeRhSn $C(T)$ data is not sufficiently well subtracted, which also leads to similar abnormal $C(T)$ behaviors at $\sim 230$ K and $\sim 289$ K, associated with the glass transition, previously reported for this contact agent \cite{Bunting1969,Schnelle1999}.
Due to such ambiguity, we do not interpret this $C(T)$ anomaly at $\sim 240$ K, but we analyze a $\Delta C \equiv C(\mathrm{CeRhSn}) - C(\mathrm{LaRhSn})$ increment, that appears for $T>100$ K, as shown in Fig. \ref{fig:Fig_C-CeRhSn}.
The maximum value of $\Delta C\sim 7$ J/(mol K) is reached at $T=T_G$ and with a further increase in $T$ the value of $\Delta C$ remains almost unchanged. In our understanding, this $\Delta C(T)$ behavior better documents the formation of the GP state below $T_G$. 

The inset in Fig. \ref{fig:Fig_C-CeRhSn} shows the behavior of $C(T)/T\propto T^{-n}$ with the exponent $n=0.4$. 
The $C/T$ vs. $T$ data also show a feature at $5.5$ K, which  is, however,  not typical for the glassy transition.
Similarly, ac susceptibility does not show a characteristic frequency dependence at this temperature, although $\chi{''}$ shows a broad peak located around 6 K (see Fig. \ref{fig:CeRhSn_CHI_ac}).
This behavior at $\sim 6$~K is more suggestive of a transition between the classical and quantum
GP states.
LaRhSn behaves differently; no anomalies were detected in its specific heat.
In the high temperature limit $C$ of LaRhSn reaches the value of $3NR =74.8$ J(mol K)$^{-1}$, which is consistent with the Dulong-Petit law ($R$ is the gas constant and $N$ is the number of atoms in the formula unit), as shown in Fig. \ref{fig:Fig_C-CeRhSn}.

\begin{figure}[h!]
\includegraphics[width=0.42\textwidth]{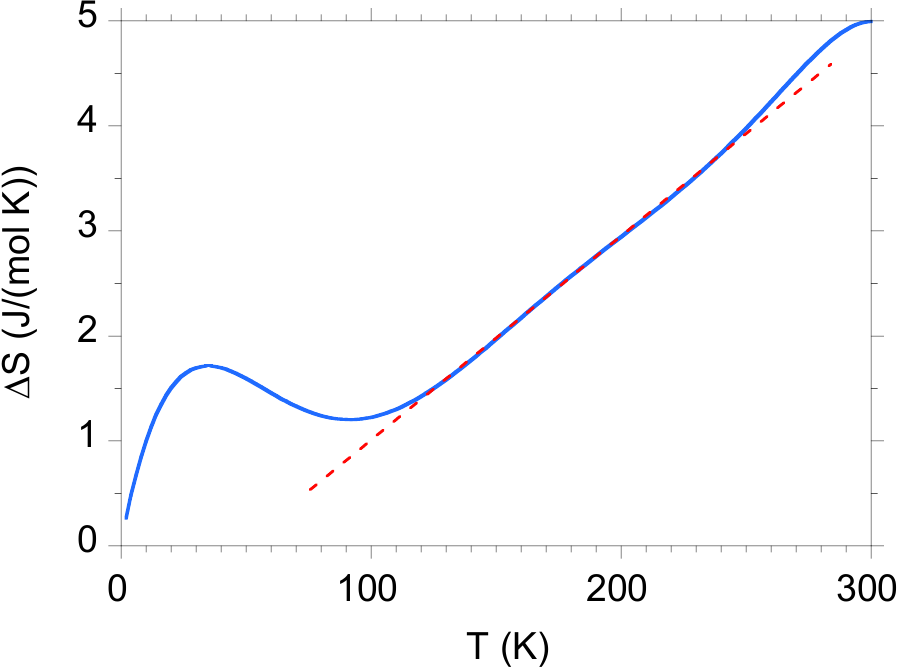}
\caption{\label{fig:DELTA_S}
Temperature variation of magnetic entropy $\Delta S=S(\mathrm{CeRhSn}) - S(\mathrm{LaRhSn})$ of CeRhSn after subtracting the phonon contribution. The dashed line (red) shows the linear dependence of $\Delta S(T)$.
}
\end{figure}

Figure \ref{fig:DELTA_S}~shows magnetic contribution $\Delta S=S(\mathrm{CeRhSn}) - S(\mathrm{LaRhSn})$ to the entropy $S$ of CeRhSn.
For temperatures $T<300$ K, $\Delta S(T)$ does not reach the value $R\ln 2 = 5.76$  J/(mol K) that corresponds to the doublet ground state of Ce and reveals a linear change with $T$ for $\sim 100<T<235$ K, i.e., within the temperature range, where a limited number of small and noninteracting spin clusters appear.  A broad maximum in $\Delta S(T)$ at $\sim 30$ K could be related to the formation of the CG state. It could also be related to the strongly frustrated arrangements of $4f$ moments in CeRhSn, as previously reported by Tokiwa {\it et al.} \cite{Tokiwa2015}. 
For comparison, the inset of Fig. \ref{fig:Fig_C-CeRhSn} displays a broad peak in the $\Delta (C/T)=C/T(\mathrm{CeRhSn}) - C/T(\mathrm{LaRhSn})$ vs $T$ presentation with a maximum in $\Delta (C/T)$ at $T\sim 20$ K, which also signals an effect of inhomogeneous CG-like magnetic order.
The thermodynamic properties of CeRhSn at both high and low temperatures are analogous to those observed for a number of Fe-based {\it paramagnetic} Heusler alloys, with a well-documented Griffiths phase state (cf. Refs. \cite{Slebarski2011,Naka2016,Slebarski2024}).
{\it Ab initio} calculations always indicated the presence of magnetic moments on the {\it wrong-site} iron atoms due to local lattice disorder, whereas the ordered system was calculated paramagnetic. Similar band structure calculations also predict the magnetic moment on the Ce ion for CeRhSn in the presence of its local disorder.

\subsection{Toy model}\label{sec:model}
The complexity of the low-temperature behavior is likely a consequence of the interplay between various factors, including disorder, frustration, and quantum fluctuations. However, the pronounced increase in $\chi(T)$ just below $T_G$ may be attributed to the formation of a limited number of small spin clusters, a phenomenon that can be effectively captured by a simple theoretical model. We assume that the sample contains a fixed number of magnetic moments, which does not depend on the temperature, but these moments may form ferromagnetic clusters. To keep the discussion general, we do not specify whether the moments are on Ce or Rh atoms. For the purpose of a quantitative analysis, it is assumed that there are $N$ magnetic moments with a magnetic moment magnitude $s$ present in the sample. They can be just electron spins $s=\frac{1}{2}$, but they can also be higher moments. If they form a cluster consisting of $i$ ferromagnetically aligned moments, the total moment of the cluster would be $J=is$. The dependence of magnetization $M$ of the ideal paramagnet on magnetic field $B$ is given by the Brillouin function
\begin{equation}
    M=Ng\mu_{\rm B} J B_J(x),
    \label{eq:magnet}
\end{equation}
where
\begin{equation}
    x=\dfrac{g\mu_{\rm B}B}{k_{\rm B}T}.
\end{equation}
For temperatures of the order of $T_G$ the argument of the Brillouin function $x\ll 1$ and in this limit Eq.~\eqref{eq:magnet} reduces to the Curie law
\begin{equation}
    M=\dfrac{Ng^2\mu_{\rm B}^2J(J+1)B}{3k_{\rm B}T},
\end{equation}
which can be written as
\begin{equation}
    M=AJ(J+1)\dfrac{B}{T}.
    \label{eq:simple_magnet}
\end{equation}
Then, the susceptibility
\begin{equation}
    \chi=\dfrac{M}{B}=A\dfrac{J(J+1)}{T}.
    \label{eq:chi_T0}
\end{equation}
When the temperature is reduced below approximately $T_0=230$~K clusters are formed. 
We assume that for $T<T_0$ we have $n(J)$ clusters with magnetic moment $J$. Since the total number of elementary moments $s$ is equal to $N$, we have
\begin{equation}
    N=\sum_{i=1}^{\infty} n(is)i
    \label{eq:total_num}
\end{equation}
We assume a simple distribution of cluster sizes \cite{PLee},
\begin{equation}
    n(J)\propto J^{-a},
    \label{eq:distrib}
\end{equation}
where the exponent is temperature dependent $a=a(T)$. Then,  
\begin{equation}
    n(J)=\kappa J^{-a},
\end{equation}
and Eq.~\eqref{eq:total_num} can be rewritten as
\begin{equation}
    N=\kappa \sum_{i=1}^{\infty}(is)^{-a}i. 
    \label{eq:Neq}
\end{equation}
If $a>2$, the above can be written as
\begin{equation}
    N=\kappa s^{-a}\zeta(a-1),
\end{equation}
where $\zeta(\ldots)$ is the Riemann zeta function. This equation gives the normalization constant $\kappa$. 
In the case where $a<2$, the series in Eq.~\eqref{eq:Neq} is not convergent. However, for temperatures exceeding $T_0$, the magnetic moments are not clustered, resulting in a strongly peaked distribution \eqref{eq:distrib} at the lowest possible value of $J$, which is $s$. This corresponds to $a\to\infty$, which allows us to assume that at temperatures slightly below $T_0$, the condition $a > 2$ is still satisfied. Subsequently, according to Eq. \eqref{eq:simple_magnet}, the total susceptibility originating from the clusters can be expressed as
\begin{align}
    \chi&=\frac{A}{NT}\sum_J n(J)J(J+1) \nonumber \\
    &=\frac{As}{T}\left[s\frac{\zeta(a-2)}{\zeta(a-1)}+1\right].
    \label{eq:clusters}
\end{align}
The above is valid only for $a>3$, but below we show that this condition is always satisfied. 

At $T_0$ we have no clusters, so $J=s$ and Eq.~\eqref{eq:chi_T0} gives
\begin{equation}
    \chi(T_0)=\dfrac{A}{T_0}s(s+1).
    \label{eq:chi_T0a}
\end{equation}
Combining Eqs.~\eqref{eq:clusters} and \eqref{eq:chi_T0a} we can write
\begin{equation}
    \dfrac{\chi(T<T_0)}{\chi(T_0)}=\left[s\frac{\zeta(a-2)}{\zeta(a-1)}+1\right]\frac{1}{s+1}\left(\dfrac{T}{T_0}\right).
    \label{eq:chi_ratio}
\end{equation}
Since $\zeta(a-2)/\zeta(a-1)$ diverges as $a\to 3$ from above, for arbitrarily strong increase of the susceptibility the exponent $a$ in Eq.~\eqref{eq:distrib} will never drop to 3 or below. The distribution for $\infty > a > 3$ is rapidly decaying, which means that even an admixture of a small number of small clusters can significantly increase susceptibility.

By measuring the ratio on the left-hand side of Eq.~\eqref{eq:chi_ratio} one can determine the exponent $a(T)$, which allows calculating the temperature dependence of the average number of elementary moments in a cluster as the ratio of the average cluster moment to $s$, that is,
\begin{equation}
    \bar{n}(T)=\dfrac{\displaystyle\sum_{J}n(J)J}{s\displaystyle\sum_{J}n(s)}=\dfrac{\zeta(a-1)}{\zeta(a)}.
    \label{eq:av_cluster}
\end{equation}

\begin{figure}[h!]
\includegraphics[width=0.45\textwidth]{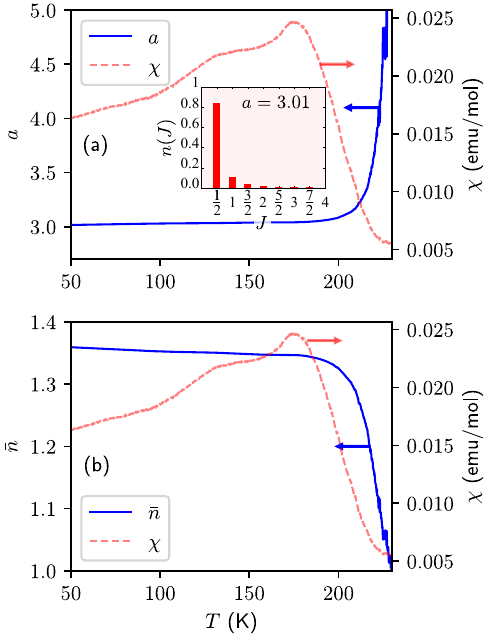}
\caption{(a) The blue line shows the temperature dependence of the exponent $a$ [see Eq.~\eqref{eq:distrib}] for the magnetic filed $B=0.005$ T and for the elementary moment $s=\frac{1}{2}$. The dashed red line shows the susceptibility $\chi(T<T_0)$ that enters Eq.~\eqref{eq:chi_ratio}, which was used to calculate $a$. The inset illustrates the distribution for ${a=3.01}$. (b) The blue line shows the temperature dependence of the size of an average cluster $\bar{n}$ [see Eq.~\eqref{eq:av_cluster}] for the same parameters as in panel (a).
\label{fig:toy_model}}
\end{figure}  

In order to demonstrate the predictions of the model, we applied it to the case where $s=\frac{1}{2}$, and $\chi(T)$ is given by the red line in Fig.~\ref{fig:CeRhSn_CHI_ac} ($B=0.005$ T). The results are presented in Fig.~\ref{fig:toy_model}. At $T=T_0$, which we assumed to be 230 K, there are no clusters, so $a=\infty$ and $\bar{n}=1$. Then, with decreasing temperature, even a small increase in susceptibility requires a rapid drop in $a$. However, both $a$ and $\bar{n}$ quickly saturate at $a\gtrapprox 3$ and $\bar{n}\approx 1.35$. This means that even the large increase in $\chi$ around $T=180$ K can result from a small admixture of small spin clusters. The size distribution presented in the inset in Fig.~\ref{fig:toy_model}(a) shows that the concentration of clusters decreases rapidly as their size increases, while single unclustered moments continue to be predominant. 

The restriction $a>3$ seems somewhat artificial, but is due to the assumption of a particular form of the cluster size distribution \eqref{eq:distrib}. Nevertheless, the conclusion that such a strong increase in susceptibility may be due to the appearance of very small and dilute clusters can also be expected to remain true for more general distributions.

\section{Concluding remarks\label{sec:conclusions}}
\begin{figure}[h!]
\includegraphics[width=0.45\textwidth]{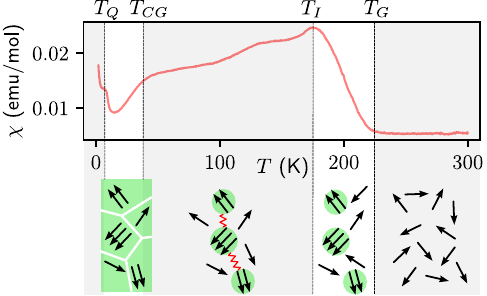}
\caption{Interpretation of the underlying microscopic states responsible for the susceptibility: for temperatures above $T_G$ the system includes single magnetic moments, at $T_G$ noninteracting clusters starts to form, resulting in a dramatic increase of $\chi$, as described in Sec.~\ref{sec:model}; at $T_I$ the intercluster interactions begin to overcome thermal fluctuations thereby gradually reducing $\chi$. At $T_{CG}$ they lead to freezing of the clusters and the formation of cluster spin glass. With a further lowering of the temperature, below $T_Q$, a quantum Griffiths phase gives rise to the power-law singularities, which is modified at much lower temperatures by possible metamagnetic transition, as has been suggested in Ref. \cite{Tokiwa2015}.
Note that these graphics only schematically illustrate the magnetic structure; in reality, the clusters are much more diluted. 
\label{fig:summary}}
\end{figure}

Comprehensive studies of CeRhSn carried out in the temperature range $T<T_G\sim 220$ K clearly indicate the presence of the Griffiths phases and also correlate well with the theoretical approach proposed by Vojta \cite{Vojta2010}. 
We focused here on the interpretation of the results obtained in the temperature range $T>2$ K; however, our research does not exclude the previously postulated critical behaviors resulting from the frustration effect \cite{Tokiwa2015} in the mK area.
The experimental results presented here allowed us to attempt to propose a microscopic interpretation of the unique behaviors observed. 
From the characteristic points in $\chi(T)$ it is possible to indicate some temperatures within the Griffiths phase at which the magnetic structure of the compound changes qualitatively. In particular, we relate these temperatures to the nonmonotonic character of $\chi(T)$. Figure~\ref{fig:summary} summarizes one of the possible scenarios. The system shows two regions of rapid increase in susceptibility, the first between $T_G$ and $T_I$ and the second at $T_Q$. While the former has been explained in Sec.~\ref{sec:model}, we relate the latter to the complex physics that emerges when classical effects begin to interact or compete with quantum effects, as mentioned in the text. The observed decrease in $\chi(T)$ between $T_I$ and $T_Q$, and particularly between $T_{CG}$ and $T_Q$, is attributed to the interactions between the clusters and their surrounding environment, as well as between the clusters themselves. These interactions are assumed to be responsible for the suppression of cluster dynamics.

The Griffiths phase scenario therefore seems to be justified for $T<T_G$. It can be observed from the dc $1/\chi$ vs $T$ plots that the GP is sensitive to magnetic fields, as the external field increases.
This phenomenon gradually weakens, and it is observed to disappear at the applied fields larger than 1~T. 
Analogous behavior in $\chi^{-1}(T)$ with increasing of $B$ was observed for other systems where the Griffiths phase is already well documented.  
The results of ac magnetization show that the GP interplays with the SG below $T_{CG}$; similar behavior was reported very recently for Ni$_2$MnSb polycrystalline Heusler alloy \cite{Tian2024}. Following Vojta \cite{Vojta2010} we attributed this diluted cluster glass phase to the presence of {\it rare regions}, where the slow dynamics of clusters can give rise to the power-law singularities of $\chi$. Our modeling does not exclude the possible metamagnetic crossover in CeRhSn at the lowest temperatures, which could lead to frustration-induced quantum criticality, as proposed by Tokiwa {\it et al.} \cite{Tokiwa2015}.

$^{\star}$Author to whom correspondence should be addressed: andrzej.slebarski@us.edu.pl

\begin{acknowledgments}
Numerical calculations have been carried out using high-performance computing resources provided by the Wroc{\l}aw Centre for Networking and Supercomputing. The authors thank M. Fija\l kowski and K. Szot for a valuable discussion.
\end{acknowledgments}


\begin{thebibliography}{50}

\bibitem{Anderson1958}
P. W. Anderson, Absence of diffusion in certain random lattices, Phys. Rev. {\bf 109}, 1492 (1958).

\bibitem{Mydosh1999}
J. A. Mydosh, Disorder and frustration in heavy-fermion compounds, Physica B {\bf 259—261}, 882 (1999).

\bibitem{Spalek2003}
J. Spa\l ek and W. W\'{o}jcik, A strong effect of disorder on Mott transition: Hubbard-Lloyd model, Acta Phys. Pol. B
{\bf 34}, 399 (2003).

\bibitem{Grenzebach2008}
C. Grenzebach, F. B. Anders, G. Czycholl, and T. Pruschke, Influence of disorder on the transport properties of heavy-fermion systems, Phys. Rev. B {\bf 77}, 115125 (2008).




\bibitem{JLTPh1996} 
Cf. special issue of J. Phys.: Conden Matter {\bf 8}, No. 48 (1996), (edited by P. Coleman, M. B. Maple, and A. Millis).

\bibitem{Gebhard1998} 
F. Gebhard, {\it The Mott Metal-Insulator Transition}, (Spronger Verlag, Berlin 1997); M. Imada, A. Fujimori, and T. Tokura, Rev. Mod. Phys. {\bf 70}, 1039 (1998).


\bibitem{Dobrosavlevic1992}
V. Dobrosavljevi\'c, T. R. Kirkpatrick, and G. Kotliar, Kondo effect in disordered systems, Phys. Rev. Lett. {\bf 69} 1113 (1992).


\bibitem{Miranda1996}
E Miranda, V Dobrosavljevi\'c, and G Kotliar, Kondo disorder: a possible route towards non-Fermi-liquid behaviour, J. Phys.: Condens. Matter {\bf 8}, 9871 (1996).

\bibitem{Miranda1997}
E. Miranda, V. Dobrosavljevi\'c, and G. Kotliar, Disorder-driven non-Fermi-liquid behavior in Kondo alloys, Phys. Rev. Lett. {\bf 78}, 290 (1997).

\bibitem{Bernal1995} 
O. O. Bernal, D. E. MacLaughlin, H. G. Lukefahr, and B. Andraka, Copper NMR and thermodynamics of UCu$_{5-x}$Pd$_x$: Evidence for Kondo disorder, Phys. Rev. Lett. {\bf 75}, 2023 (1995).

\bibitem{Andraka1993}
B. Andraka and G. R. Stewart, Heavy-non-Fermi-liquid behavior in U(Cu,Pd)$_5$, Phys. Rev. B {\bf 47}, 3208 (1993); M. C. Aronson, R. Osborn, R. A. Robinson, J. W. Lynn, R. Chau, C. L. Seaman, and M.B. Maple, Non-Fermi-liquid scaling of the magnetic response in UCu$_{5-x}$Pd$_x$ (x = 1, 1.5), Phys. Rev. Lett. {\bf 75}, 725 (1995). 

\bibitem{Seaman1993}
C. Seaman, M. B. Maple, B. W. Lee, S. Ghamaty, M. S. Torikachvili, J. -S. Kang, L. Z. Liu, J. W. Allen, and D. L. Cox, Evidence for non-Fermi liquid behavior in the Kondo alloy Y$_{1-x}$U$_x$Pd$_3$, Phys. Rev. Lett. {\bf 67}, 2882 (1991); B. Andraka and A. M. Tsvelik, Observation of non-Fermi-liquid behavior in U$_{0.2}$Y$_{0.8}$Pd$_3$, Phys. Rev. Lett. {\bf 67}, 2886 (1991).

\bibitem{Maple1995}
M. B. Maple, M. C. de Andrade, J. Herrmann, Y. Dalichaouch, D. A. Gajewski, C. L. Seaman, R. Chau, R. Movshovieh, M. C. Aronson, and R. Osborn, Non Fermi liquid ground states in strongly correlated f-electron materials, J.Low Temp. Phys. {\bf 99}, 223 (1995).

\bibitem{Andraka1994a}
B.Andraka, Anomalous low-temperature properties of dilute Ce alloys Ce$_x$La$_{1-x}$Cu$_{2.2}$Si$_2$ ($x\leq 0.2$), Phys. Rev. B {\bf 49}, 3589 (1994). 
  
\bibitem{Andraka1994b}
B. Andraka, Anomalous specific heat of Ce$_{1-x}$Th$_x$RhSb alloys, Phys. Rev.B {\bf 49}, 348 (1994). 

\bibitem{Castro1998}
A. H. Castro Neto, G. Castilla, and B. A. Jones, Non-Fermi liquid behavior and Griffiths phase in $f$-lectron compounds, Phys. Rev. Lett. {\bf 81}, 3531 (1998). 

\bibitem{Castro2000}
A. H. Castro Neto and B. A. Jones, Non-Fermi-liquid behavior in U and Ce alloys: Criticality, disorder, dissipation, and Griffiths-McCoy singularities, Phys. Rev. B {\bf 62}, 14975 (2000).

\bibitem{Andrade1998}
M. C. de Andrade, R. Chau, R. P. Dickey, N. R. Dilley, E. J. Freeman, D. A. Gajewski, M. B. Maple, R. Movshovich, A. H. Castro Neto, G. Castilla, and B. A. Jones, Evidence for a common physical description of non-Fermi-liquid behavior in chemically substituted f-electron systems, Phys. Rev. Lett. {\bf 81}, 5620 (1998).

\bibitem{Volmer2000}
R. Vollmer, T. Pietrus, H. v. L\"{o}hneysen, R. Chau, and M. B. Maple, Phase transitions and non-Fermi-liquid behavior in UCu$_{5-x}$Pd$_x$ at low temperatures, Phys. Rev. B {\bf 61}, 1218 (2000).

\bibitem{Griffiths1961}
R. B. Griffiths, Nonanalytic behavior above the critical point in a random Ising ferromagnet, Phys. Rev. Lett. {\bf 23}, 17 (1969).

\bibitem{comment1}
Griffiths discussed a nonanalytic behavior of the magnetization above Curie temperature  in a randomly diluted Ising ferromagnet, caused by the formation of ferromagnetic clusters  \cite{Griffiths1961}. The intermediate phase located between the ferromagnetic and paramagnetic state of the sample is referred to as the Griffiths phase.


\bibitem{Vojta2010}
T. Vojta, Quantum Griffiths effects and smeared phase transitions in metals: Theory and experiment, J. Low. Temp. Phys. {\bf 161}, 299 (2010).


\bibitem{Slebarski2002a}
A. \'{S}lebarski, M. B. Maple, E. J. Freeman, C. Sirvent, M. Rad\l owska, A. Jezierski, E. Granado, Q. Huang, and J. W. Lynn, Strongly correlated electron behaviour in the compound CeRhSn, Phil. Mag. B {\bf 82}, 943 (2002).

\bibitem{Slebarski2002b}
A. \'{S}lebarski, N. A. Frederick, and M. B. Maple, Strongly correlated electron behaviour in stoichiometric CeRhSn and non-stoichiometric Ce$_x$RhSn, Phil. Mag. B {\bf 82}, 1275 (2002).



\bibitem{Ho2004}
P. -C. Ho, V. S. Zapf, A. \'Slebarski, and M. B. Maple, Non-Fermi-liquid behavior in CeRhSn, Phil. Mag. {\bf 84}, 2119 (2004).






\bibitem{Kim2003}
M. S. Kim, Y. Echizen, K. Umeo, S. Kobayashi, M. Sera, P. S. Salamakha, O. L. Sologub, T. Takabatake, X. Chen, T. Tayama, T. Sakakibara, M. H. Jung, and M. B. Maple, Low-temperature anomalies in magnetic, transport, and thermal properties of single-crystal CeRhSn with valence fluctuations, Phys. Rev. B {\bf 68}, 054416 (2003).

\bibitem{Tou2004}
H. Tou, M. S. Kim, T. Takabatake, and M. Sera, Antiferromagnetic spin fluctuations in CeRhSn probed by $^{119}$Sn NMR,  Phys. Rev. B {\bf 70}, 100407(R) (2004). 

\bibitem{Tokiwa2015}
Y. Tokiwa, Ch. Stingl, M. -S. Kim, T. Takabatake, and P. Gegenwart, Characteristic signatures of quantum criticality driven by geometrical frustration, Sci. Adv. {\bf 1}, e1500001 (2015).

\bibitem{Kittaka2021}
S. Kittaka, Y. Kono, S. Tsuda, T. Takabatake, and T. Sakakibara, Field-angle-resolved landscape of non-Fermi-liquid behavior in the quasi-Kagome Kondo lattice CeRhSn, J. Phys. Soc. Jpn. {\bf 90}, 064703 (2021).

\bibitem{Kimura2024}
Shin-ichi Kimura, M. F. Lubis, H. Watanabe, Y. Shimura, and T. Takabatake, Anisotropic non-Fermi liquid and dynamical Planckian scaling of the quasi-kagome Kondo lattice CeRhSn, 
https://doi.org/10.48550/arXiv.2402.18176.

\bibitem{Bohm2024}
T. U. B\"{o}hm, N. S. Sirica, B. G. Jang, Y. Liu, E. D. Bauer, Y. Huang, Ch. C. Homes, J. -X. Zhu, and F. Ronning, Anisotropic hybridization in CeRhSn, Phys. Rev. B {\bf 110}, L121107 (2024).


\bibitem{Slebarski2004} 
A. \'Slebarski, K. Grube, R. Lortz, C. Meingast, and H. v. L\"{o}hneysen, Thermal and magnetic properties of CeRhSn, J. Magn. Magn. Mater. {\bf 272-276}, 234 (2004).

\bibitem{Slebarski2011a}
A. \'Slebarski, Gr\"{u}neisen ratio in Kondo-lattice compound CeRhSn, J. Phys.: Conf. Ser. {\bf 303}, 012109 (2011).

\bibitem{Zhu2003}
L. Zhu, M. Garst, A. Rosch, and Q. Si, Universally diverging Grüneisen parameter and the magnetocaloric effect close to quantum critical points, Phys. Rev. Lett. {\bf 91}, 066404 (2003).

\bibitem{comment_QCP}
Anisotropy in the linear thermal expansion has been found in number of quantum critical tetragonal or orthorhombic heavy fermion metals (cf. references in \cite{Tokiwa2015}), however, in these materials, divergent behavior in $\alpha/T$ can be observed along all main directions. This behavior is different in CeRhSn.  

\bibitem{Rodriguez1993}
J. Rodriguez-Carvajal, Recent advances in magnetic structure determination by neutron powder diffraction, Physica B {\bf 192}, 55 (1993).

\bibitem{Toby2006}
B. H. Toby, R factors in Rietveld analysis: How good is good enough? Powder Diffr. {}
{\bf 21}, 67 (2006).


\bibitem{Apiezon}
Apiezon Products, M\&I Materials, Manchester, M32 0ZD
United Kingdom

\bibitem{Schnelle1999}
W. Schnelle, J. Engelhardt, and E. Gmelin, Specific heat capacity of Apiezon N high vacuum grease and of Duran borosilicate glass, Cryogenics {\bf 39}, 271 (1999).

\bibitem{Hill1970}
H. H. Hill, The Early Actinides: the Periodic System’s f Electron Transition Metal Series, in Plutonium 1970 and Other Actinides (AIME, New York, 1970).




\bibitem{Slebarski2003}
A. \'Slebarski and A. Jezierski,  Non-Fermi liquid behavior in CeRhSn coexistent with magnetic order, Phys. Stat. Sol. B {\bf 236}, 340 (2003).

\bibitem{commentK}
In CeRhSn, Kondo temperature is about 140 K \cite{Slebarski2002a}


\bibitem{Cox1994}
A. J. Cox, J. G. Louderback, S. E. Apsel, and L. A. Bloomfield, Magnetism in $4d$-transition metal clusters, Phys. Rev. B {\bf 49}, 12295 (1994).

\bibitem{Costro1997}
M. Castro, C. Jamorski, and D. R. Salahub, Structure, bonding, and magnetism of small Fe$_n$, Co$_n$, and Ni$_n$ clusters, $n \leq 5$, Chem. Phys. Lett. {\bf 271}, 133 (1997).

\bibitem{Jo2006}
Y. Jo, M. H. Jung, M. C. Kyum, K. H. Park, and Y. N. Kim, Magnetic properties of nano-sized CuNi clusters, J. Magnetics, {\bf 11}, 156 (2006).

\bibitem{Slebarski2002c}
A. \'Slebarski, M. Rad\l owska, T. Zawada, M. B. Maple, A. Jezierski, and A. Zygmunt, Experimental study of the physical properties in the complex magnetic phase diagram of Ce$_{1-x}$La$_x$RhSn, Phys. Rev. B {\bf 66}, 104434 (2002).


\bibitem{Bunting1969}
J. G. Bunting, T. Ashworth, and H. Steeple, A correlation between thermal conductance and specific heat anomalies and the glass temperature of Apiezon N and T greases, Cryogenics {\bf 9}, 385 (1969).

\bibitem{Slebarski2011}
A. \'{S}lebarski, J. Goraus, and M. Fija\l kowski, Short-range ferromagnetic correlations in disordered FeVGa with distinct similarities to the Griffiths phase, Phys. Rev. B {\bf 84}, 075154 (2011).

\bibitem{Naka2016}
T. Naka, A. M. Nikitin, Yu Pan, A. de Visser, T. Nakane, F. Ishikawa, Y. Yamada, M. Imai, and A. Matsushita, Composition induced metal–insulator quantum phase transition in the Heusler type Fe$_2$VAl, J. Phys.: Condens. Matter {\bf 28}, 285601 (2016).

\bibitem{Slebarski2024}
A. \'{S}lebarski, M. Fija\l kowski, J. Deniszczyk, M. M. Ma\'{s}ka, and D. Kaczorowski, Off-stoichiometric effect on magnetic and electron transport properties of Fe$_2$VAl$_{1.35}$ and Ni$_2$VAl: A comparative study, Phys. Rev. B {\bf 109}, 165105 (2024).


\bibitem{PLee}
B. S. Shivaram, J. C. Prestigiacomo, A. Xu, Zhenyuan Zeng, T. D. Ford, I. Kimchi, S. Li, and P. A. Lee, Nonanalytic magnetic response and intrinsic ferromagnetic clusters in a kagome spin-liquid candidate, Phys. Rev. B {\bf 110}, L121105 (2024).

\bibitem{Tian2024}
F. Tian, Q. Zhao, J. Guo, Y.  Zhang, M. Fang, T. Chang, Z. Dai, Ch. Zhou, K. Cao, and S. Yang,
Griffiths phase arising from local lattice distortion and spin glass above the Curie
temperature in Ni$_2$MnSb polycrystalline Heusler alloy, Phys. Rev. B {\bf 109}, 224405 (2024).






\end{thebibliography}
\end{document}